\documentclass[12pt]{article}

\usepackage{amsmath}
\usepackage{amssymb}

\begin{document}

\ \vskip 1.0 in

\begin{center}
 { \large {\bf The Three and a Half Layers of Dynamics  }}

\smallskip

{ {\bf   {\it Analog, Digital, Semi-digital, Analog} }}

\vskip 0.2 in

\smallskip

\bigskip

\bigskip

\bigskip

{\large{\bf Tejinder Singh}}

\medskip

{\it Tata Institute of Fundamental Research,}\\
{\it Homi Bhabha Road, Mumbai 400 005, India.}\\
{\tt email: tpsingh@tifr.res.in}\\
\medskip

\vskip 0.5cm
\end{center}

\vskip 1.0 in

\begin{abstract}

\noindent Quantum theory is extremely successful in explaining most physical phenomena, and is not contradicted by any experiment. Yet, the theory has many puzzling features : the occurrence of probabilities, the unclear distinction between the microscopic and the macroscopic, the unexplained absence of superpositions in positions of macroscopic objects, the dependence of the theory on an external classical time, and the experimentally verified but peculiar `influence' outside the light-cone in EPR experiments. These puzzles point towards a conflict between quantum theory and our present understanding of spacetime structure, and suggest the existence of a deeper theory. In this essay we make the case that in the underlying theory the matter and spacetime degrees of freedom are non-commuting matrices, and yet the dynamics is analog. A digital quantum-theory like dynamics for matter as well as spacetime emerges in the statistical thermodynamic approximation to this deeper theory. When most of the matter clumps into macroscopic structures, it is shown to behave classically, and it induces classical dynamics on spacetime; this is the eventual analog limit, our macroscopic world. In between the digital layer and the uppermost analog layer is the realm of standard quantum theory - microscopic objects and their interaction with measuring apparatuses on a classical spacetime background : the semi-digital approximation. Such a multi-layered description of dynamics can explain the puzzling features of quantum theory, and is testable by ongoing laboratory experiments.

\end{abstract}

\vskip 0.5 in

\centerline{\it This essay received the fourth prize in the Essay Competition (2010/11) "Is reality digital or analog?" }

\centerline{conducted by the Foundational Questions Institute. http://fqxi.org/community/essay/winners/2011.1}

\newpage


\noindent The need to explain the spectrum of black-body radiation, the photo-electric effect, and the spectral lines of atoms, compelled physicists to give up the classical Newtonian picture of dynamics, and led to the development of the formalism of quantum theory in the 1920s. The dynamics of particles is encoded in the wave-function, which evolves according to the Schr\"{o}dinger equation, or, in the relativistic case, the Dirac equation. Position and momenta no longer commute, and their non-commutativity is expressed by a universal relation involving the Planck constant, thus giving the theory its digital structure.  Besides explaining the aforementioned observations, the theory was spectacularly successful in aiding the development of a relativistic quantum field theory of elementary particles. Its predictions in quantum electrodynamics, the electro-weak theory, and quantum chromodynamics, agree very well with laboratory experiments. The `action-at-a-distance' in an EPR type experiment, which Einstein called spooky, has been experimentally proven to exist [however, we know it cannot be used for signalling]. On the technological front, quantum theory has yielded immensely useful applications such as semiconductors and lasers, and is at the basis of exciting new fields such as nanotechnology, quantum computing, quantum information and quantum cryptography. Above all, there is no experiment to date which contradicts the predictions of the theory.    

And yet, no avid student of quantum mechanics is left untouched by some of the puzzling and discomforting features of the theory. Making the plausible demand that quantum mechanics should describe individual systems and not their statistical ensembles, foremost amongst such discomforting features is the unexpected occurrence of probabilities in an otherwise dynamical theory. If a classical apparatus is used to make a measurement  on a quantum system which is initially in a state which is a superposition of the eigenstates of the observable being measured, the outcome cannot be predicted with certainty. Only the probability of occurrence of a resulting eigenstate is known, and this is equal to the square of the amplitude for the initial state to be in the said eigenstate [the Born probability rule]. Why should probabilities arise when the initial conditions [the initial quantum state, and the amplitudes to be in various eigenstates] are precisely known? Contrast this with the tossing of a fair dice, where we know that the outcome cannot be predicted with certainty because of our ignorance of initial conditions. Moreover, unlike in the case of tossing of the dice, where the probability of any one outcome can be reasoned to be 1/6, quantum theory provides no explanation for the Born rule, simply adopting it as an ad hoc postulate which agrees well with experiment.   

The next puzzling aspect is the ill-defined distinction between the microscopic quantum system, and the macroscopic classical apparatus. After all, the classical apparatus is nothing but a special quantum system, and from this viewpoint one is faced with the following question : why macroscopic systems (unlike microscopic ones) are not found in superposition of position states [the Schr\"{o}dinger cat paradox]? This, in spite of the fact that the principle of linear superposition is a hallmark of quantum theory. How large should a system be before it ceases to be a quantum system and becomes classical? Quantum theory does not answer this question. Explanations such as those provided by non-local hidden variable theories, or the many-worlds interpretation, seem unfalsifiable, because they are reinterpretations of quantum theory which make the same experimental predictions as the standard theory.   

Thirdly, the current formulation of the theory depends crucially on the presence of an external classical world. Evolution is defined with respect to an external classical time which is part of a spacetime geometry determined by classical matter fields according to the laws of general relativity. Furthermore, to obtain a quantum theory, we have to `quantize' an already known classical theory. It is not very satisfactory if for its formulation a theory has to depend on its own special limit.  
    
These puzzling features suggest that quantum theory is not exact, but an asymptotic approximation to a deeper theory. This deeper theory must not quantize a pre-given classical Newtonian theory, but derive quantum theory from first principles, and explain the origin of probabilities, and the transition from microscopic to macroscopic, without having to a priori depend on an external classical time. Very impressive progress towards developing such an underlying theory, christened Trace Dynamics, has been made by Stephen Adler and collaborators \cite{adler}. This theory does not start from a pre-existing classical mechanics of point particles; however, it takes classical time as a given. Below we summarize Trace Dynamics and its achievements. Then we propose a generalization which does not take classical time as given - a generalization which presents us with the `three and a half layers of dynamics'.      

\bigskip

\bigskip

\centerline{\it Trace Dynamics}

\medskip 

\noindent
Trace Dynamics is the classical dynamics of $N\times N$ non-commuting matrices $q_r$  [equivalently operators] whose elements can either be complex numbers [in which case the matrices are called bosonic] or Grassmann numbers [fermionic matrices]. A key ingredient is the trace derivative -  the derivative with respect to an operator, of the trace of a polynomial made out of the matrices, defined by first varying, and then cyclically permuting inside the trace. The Lagrangian of the theory is the trace of a polynomial function of the matrix variables $\{q_r\}$ and their time-derivatives $\{\dot q_r\}$. Conjugate momenta $p_r$ are defined as in classical dynamics; Lagrangian dynamics is derived from an action principle, and a conserved trace Hamiltonian and Hamilton's equations are constructed. Note that although the coordinate-coordinate, coordinate-momentum and momentum-momentum commutators are in general non-zero, this is an analog [equivalently continuum or classical] theory, because these commutators take values which are arbitrary, and not fixed. The lesson is that noncommutativity need imply a digital structure only when the commutators amongst the fundamental degrees of freedom take fixed universal values. Nonetheless, as a result of possessing global unitary invariance, the theory magically admits a Poincare invariant conserved Noether charge, which is equal to the sum of the bosonic commutators $[q,p]$ minus the sum of the fermionic anticommutators $\{q,p\}$. This anti-self-adjoint matrix-valued conserved quantity, which has the dimensions of action, and which we call the Adler-Millard constant \cite{adler-millard}, plays a profound role in the emergence of quantum theory. If $q$ and $p$ were real numbers (as in ordinary classical mechanics) instead of being matrices, this conserved quantity will be trivially zero - hence the novelty of the matrix classical dynamics.       

Since one is not interested in examining the dynamics at this level of precision, a statistical mechanics of this classical dynamics is developed, and a probability distribution derived by maximizing the entropy of the canonical ensemble, subject to the constancy of the conserved quantities. 
Assuming that the ensemble does not prefer any one state in the Hilbert space over the other, it can be shown that the canonical average of the Adler-Millard constant can be written as a real number $D$ times a diagonal, traceless, anti-self-adjoint matrix $i_{eff}$ whose elements are $\sqrt{-1}$ or $-\sqrt{-1}$. It turns out that this probability distribution is invariant under only that subset of global unitary transformations which commute with $i_{eff}$. This subset provides the largest symmetry group which allows for a non-zero value of the averaged A-M constant. Hence, in all subsequent considerations, one deals only with the projections $x_{eff}$ of dynamical variables $x$ : the $x_{eff}$ commute with $i_{eff}$. A Ward identity [analogous to the equipartition theorem] is shown to hold for any polynomial function $W$, as a consequence of requiring that the integration phase space measure is invariant under a constant shift of the dynamical matrix variables. The identity simplifies significantly under the following two assumptions : (i) the fundamental energy scale in the problem is Planck scale, and we are for now interested only in examining the dynamics at much lower energy scales, (ii) the A-M constant appearing in the identity is replaced by its canonical average. Now, if the polynomial $W$ is chosen to be the Hamiltonian, the Ward identity implies that the canonical averages of the effective dynamical variables obey the standard Heisenberg equations of motion of quantum dynamics. If $W$ is chosen to be a dynamical variable $x$ itself, the same Ward identity implies the standard commutation/anti-commutation relations of quantum theory, {\it provided} we identify the coefficient $D$ appearing in the A-M constant with Planck's constant $\hbar$. The transition from the Heisenberg picture to the Schr\"{o}dinger picture is made as in ordinary quantum theory. Thus, with a few assumptions (some plausible, and some that will require a better understanding of Trace Dynamics) one finds quantum dynamics emerging from an underlying classical matrix theory which possesses a global unitary invariance. The digital has emerged from the analog. And this has happened because in the thermodynamic approximation the Adler-Millard constant has been equipartitioned over all the degrees of freedom.

Where there is statistical thermodynamics, there are fluctuations. The next step in the scheme is to consider statistical fluctuations around the thermodynamic approximation made above. This involves giving up the assumption that in the Ward identity the Adler-Millard constant be replaced by its average; instead the constant is replaced by its average plus a term which represents Brownian motion type fluctuations. Instead of the standard linear Schr\"{o}dinger equation of quantum theory, one now gets a stochastic non-linear Schr\"{o}dinger equation; the stochastic part is contained in the nonlinear terms which have resulted on account of including the Brownian fluctuations. Two significant results emerge. If one sets up a quantum system in an initial state which is a superposition of the eigenstates of an observable $A$ which commutes with the Hamiltonian, the stochastic average of the variance of $A$ goes to zero as the time elapsed goes to infinity. This means that the system has been driven to a definite eigenstate of $A$, though we cannot predict which one. But we can calculate the probability of every outcome. This is possible because the stochastic average of the expectation value of the projectors $\pi=|a><a|$ of $A$ is a constant in time. This quantity is initially equal to 
$|<\psi|a>|^{2}$, the square of the amplitude for the system to be initially in the state $|a>$. The final value of this stochastic average, as $t\rightarrow \infty$, is the probability $P_a$ for the system to be driven to the state $|a>$. Hence $P_a = |<\psi|a>|^{2}$, which is the Born probability rule, now derived from first principles, and no longer ad hoc! The time-scale over which the system is driven to one of the eigenstates can be calculated, and depends on the number of degrees in the system - it is astronomically large for microscopic systems, and unmeasurably small for macroscopic ones. In other words, the nonlinear stochastic fluctuations are significant for macroscopic systems, but utterly negligible for microscopic ones. This is how we are to understand the dynamical collapse of the wave-function during a quantum measurement, and the absence of superpositions in large systems. The principle of linear superposition is an approximate principle of nature. The difference between micro and macro is one of degree, not of kind - the cat is indeed both alive and dead, but for an extremely small, unmeasurable time interval. It follows that if we examine mesoscopic systems [not too small, not too large - say a micro-mirror with $10^{15}$ atoms] in the laboratory, we might be able to experimentally detect the decay of carefully prepared linear superpositions, thus providing possible evidence for the theory of Trace Dynamics. More on this later.       

An exhaustive phenomenological study of dynamical collapse models has been carried out by Ghirardi, Pearle, Rimini and Weber \cite{grw}, \cite{pearle}, \cite{bassi}. The proposal that the Born probability rule can be derived from consideration of stochastic fluctuations is originally due to Pearle \cite{pearle2}. The role of spacetime structure in explaining quantum measurement has been investigated in \cite{singh}, \cite{singhlochan}.

\newpage 

\centerline{1. {\it The Underlying Analog Layer - Matter and Spacetime as Classical Matrix Variables}}

\bigskip 

\noindent
Trace Dynamics is remarkable in its scope and achievement; yet it takes an external classical time as given. This cannot be regarded as satisfactory : let us imagine a situation in which there are no external classical matter fields, and all matter is quantum. The induced spacetime metric will exhibit quantum fluctuations. Now, according to the Einstein hole argument \cite{Einstein}, \cite{christian}, a spacetime manifold in a generally covariant theory requires a definite classical metric to exist on it, in order for us to be able to attach a physical meaning to the points of spacetime. If the metric has quantum fluctuations, the underlying spacetime manifold is destroyed, taking away from us the privilege of a pre-existing classical time. Nonetheless, one should be able to describe quantum dynamics, and hence the need for going beyond Trace Dynamics. Happily, the theory provides hints for its own extension. We shall assume that not only the matter degrees of freedom are to be described by matrix variables, but the same is true of spacetime points - the coordinates $({\bf x},t)$ become matrices too. In other words, the location of every particle is described by operators $(\hat{q}, \hat{t})$. Evolution is described with respect to an auxiliary affine parameter $\tau$ (a real number) whose interpretation will become clear subsequently. The technical endnotes illustrate the physics involved; in particular we introduce a non-commutative Minkowski spacetime in which coordinates obey arbitrary non-commutation relations \cite{lochan}. A novel aspect is the generalization of the Poincare group to a larger symmetry group. The interpretation is that the dynamics is invariant under a generalized set of Lorentz transformations when the spacetime coordinates do not commute with each other.  The external classical time has been removed by raising time to the status of an operator.

One now repeats the Trace Dynamics construction, with the important generalization that with every particle there is associated, besides the conjugate matrix pair $(q,p)$, also a conjugate pair $(E, \hat{t})$, where $E$ is the energy operator conjugate to $\hat{t}$. The Trace Lagrangian, action, Lagrange equations of motion, Trace Hamiltonian and Hamilton's equations of motion are constructed as before. The theory obeys global unitary invariance, and hence admits a generalized Adler-Millard conserved Noether charge which also includes commutators/anti-commutators of $E$ and $\hat{t}$. This generalized charge can be shown to be invariant under the generalized Poincare transformations.

This then is the underlying Analog Matrix Dynamics of continuum matter and spacetime degrees of freedom, which admits a global unitary invariance and a generalized Poincare invariance. For now, we will refrain from introducing a `curved' spacetime; the key principles can be illustrated using the non-commutative Minkowski spacetime described in the endnotes.      

\bigskip

\bigskip 

\centerline{2. {\it The Digital Layer - Matter and Spacetime as Emergent Quantum Variables}}

\bigskip

\noindent
As in Trace Dynamics, one constructs a statistical mechanics for the analog dynamics described above, and obtains an equilibrium canonical distribution by maximizing the Boltzmann entropy, subject to the constraint that certain quantities such as the Trace Hamiltonian and the generalized Adler-Millard charge are conserved. A Ward identity is derived as a result of translational invariance of the phase space measure. Approximations are made that we want to observe physics much below the fundamental (Planck) scale, and that the Adler-Millard charge be replaced by its canonical average, in the Ward identity. The emergent dynamics is a quantum-theory like description of matter and spacetime degrees of freedom. The standard commutation/anti-commutation relation between position and momenta hold, but these are now accompanied by an analogous 
commutation/anti-commutation relation between energy and operator time. In this sense, we do not have a classical spacetime background, as 
{\it all} matter and spacetime degrees of freedom have operator status. Heisenberg equations of motion hold, in which evolution is with respect to the auxiliary parameter $\tau$, and now the Trace Hamiltonian is a function also of the operator time $\hat{t}$ associated with each particle.
There is an equivalent generalised Schr\"{o}dinger picture, in which the wave-function evolves with respect to the auxiliary parameter, and is a function of operator time variables as well. We can think of this theory as an equivalent of quantum theory, in the absence of classical time.

This is the digital layer of dynamics, having emerged from analog physics because of the equipartitioning of the Adler-Millard charge. Just like the underlying analog layer, this is not the dynamical layer that we as observers have access to, because here time is not classical.  

\bigskip

\bigskip

\centerline{3. {\it The Upper Analog Layer - The Classical World}}

\bigskip

\noindent
Like in Trace Dynamics one next allows for inclusion of stochastic fluctuations of the Adler-Millard charge, in the Ward identity. This leads once again to a stochastic non-linear Schr\"{o}dinger equation, but now with some additional, dramatic, consequences. Consider a situation where matter starts forming macroscopic clumps [for instance in the very early universe, soon after a Big Bang]. The stochastic fluctuations become more and more important as the number of degrees of freedom in the clumped system is increased. These fluctuations force macroscopic objects to be localized, but now not only in space, but also in time! This means that the time operator associated with every particle has become classical (a 
c-number times a unit matrix). The localization of macroscopic objects comes hand in hand with the emergence of a classical spacetime, in accordance with the Einstein hole argument and the general theory of relativity. If, and only if, the Universe is dominated by macroscopic objects, as today's Universe is, can one also talk of the existence of a classical spacetime. When this happens, the auxiliary time $\tau$ may be identified with the the classical time emerging from matrix dynamics. The Born probability rule is still at work, randomly selecting one possible macrostate out of the many that are possible. Once the Universe has reached this classical state, it sustains itself therein, by virtue of the continual action of stochastic fluctuations on macroscopic objects, thereby permitting also the existence of a classical spacetime geometry. As noted earlier, how large an object should be before we call it macroscopic depends entirely on how small a time interval we can measure - over which time we can detect linear superpositions. If we succeed in detecting superpositions lasting for smaller and smaller times, what we label as macroscopic (i.e. classical) will become larger and larger!   

What is the status of the auxiliary parameter $\tau$ when one does not have a classical spacetime? This parameter is nothing more than a convenient trajectory label which, as we now suggest, gives the misleading impression of being time in disguise. We propose that in the underlying analog dynamics the appropriate way to think of the Universe is as a Block Universe - a solution that can be derived by extremizing the action [like in the Fermat principle based derivation of Snell's law for the complete trajectory of a refracted light-ray], and wherein the matrix model exists `eternally', without reference to a time-evolution. The auxiliary parameter only labels the `trajectory'. The Block Universe description gives way to a time-evolving universe when classical time becomes available, in the sense explained above.  Where there is no classical time, the Block Universe picture is the only fundamental one. 

The transition from the lower analog layer to the upper analog layer also helps understand the origin of the arrow of time. As we well know, the question of the origin of the arrow of time is actually better stated as : why is the initial entropy of the Universe so low? If we believe that the primordial state from which the Universe originated is described by analog matrix dynamics, we have an answer. The Universe is in one precise microstate of the matrix model, even though in our present state of understanding of the theory we do not know what this state is. The Boltzmann entropy corresponding to this one microstate is obviously zero. Only later, when progress towards the formation of macroscopic objects takes place in the very early universe, does the statistical thermodynamic description and a non-trivial entropy come into play, and the entropy then starts to rise monotonically above its initial value of zero.    

Thus our everyday familiar analog world is recovered as an approximation to an underlying analog matrix dynamics. The laws of dynamics at the two levels are the same, but the `extended objects' described fundamentally as matrices are now described as point particles. In between these two levels the laws look very different [digital, commutation relations, probabilities, ...] but only because we had not been looking at things in the natural manner in which we should have!   

\bigskip

\bigskip 

\centerline{2.5 {\it The Semi-Digital Layer : Quantum Theory on a Classical Background}}

\bigskip

\noindent 
Until the discovery of quantum theory nearly a century ago, the only layer of dynamics we were familiar with was the analog classical universe. All that changed with the experimental evidence for the quantum nature of radiation and matter, and the inevitable development of quantum theory as we know it took place. The discomforting features of the theory, described in the Introduction, compel one to look for deeper theories such as Trace Dynamics and Analog Matrix Dynamics, and these provide an entirely fresh way of looking at quantum theory.

The current Universe is dominated by macroscopic objects which give rise to a classical spacetime geometry and the notion of time. This approximate description is a consequence of the strongly acting stochastic fluctuations which cause the transition from the underlying analog layer to the upper analog layer, via the intermediate digital layer where both matter and spacetime exhibit quantum behaviour. But this digital layer is not what we see. We experience spacetime as classical. So to get to what we see in our laboratory experiments, we must first allow for the transition to the upper analog layer, so that we have in hand a classical time. 

That done, we realize that although almost all of the matter in the universe is clumped into macroscopic objects, innumerable laboratory experiments [as also astronomical observations such as stellar spectra and the black-body nature of the Cosmic Microwave Background Radiation] provide inescapable evidence for the sub-dominant microscopic world where dynamics seems to have different rules. To arrive at these, we first develop and derive the concept of a classical time, as described above. Taking this classical time as {\it given}, one constructs the Trace Dynamics of the matter degrees of freedom, thus obtaining standard quantum theory in the thermodynamic approximation, and the Born probability rule and an explanation for the measurement problem by considering Brownian fluctuations around this thermodynamic approximation. We call this the 
semi-digital layer : a classical universe and a classical time are externally given, and on this background the digital dynamics of microscopic objects can be described. In this new way of looking at things the discomforting aspects of quantum theory disappear, having been explained away with the knowledge that the theory is not exact but approximate.         

Perhaps this is a good place to mention that what is sometimes called the paradox of the wave-particle duality is no paradox at all from our 
viewpoint. In quantum theory the true nature of microscopic objects is wave-like; it is only when they interact with and become part of a localized macroscopic apparatus that we assign them point-like particle characteristics.

\newpage 

\centerline{\it Experimental Tests}

\bigskip

\noindent
It would not do to only derive standard quantum theory from something more fundamental, nor to only show that classical mechanics is a limiting case of quantum mechanics. The theoretical framework must have predictive power, and fortunately it does. Quantum theory says that the lifetime of a linear superposition of quantum states is infinite. The theory described here says otherwise : this lifetime is finite, being astronomically large for micro-systems, and unmeasurably small for macro-systems. For mesoscopic systems this lifetime is neither too large nor too small, say in the range of a nanosecond to a microsecond, and hence in principle measurable in the laboratory. An experiment which can detect the decay of an initially prepared superposition will be evidence for departure from quantum theory, and the work described here is one possible candidate for the explanation of such a departure.   

Intense experimental effort is currently underway to test quantum mechanics for mesoscopic systems. Interference experiments aim to test the validity of linear superposition for larger and larger molecular systems such as fullerenes \cite{zeilinger} having about a hundred atoms are so.  
These experiments are very impressive; yet they have some way to go before they enter the truly mesoscopic terrain of an object having say a billion atoms. [For comparison, an object with Planck mass (= $10^{-5}$ grams) has some $10^{18}$ atoms]. Another class of experiments attempts to create superposed states for mesoscopic objects by eliminating environmental decoherence via lowering the system to extremely small temperatures
\cite{bouwmeester}, \cite{vienna}, \cite{santa barbara}. One experiment seeks to create a superposition of two position states of a micro-mirror by entangling them with two different photon eigenstates [micromechanical resonator].  Another experiment seeks to create the superposition of position states of a nanomechanical resonator [a silicon whisker] by electronically entangling it with the positive and negative charge states
of a superconducting Cooper-Pair Box. As of now experiments are inconclusive as to whether or not quantum theory holds for mesoscopic systems, the greatest technological challenge being to eliminate environmentally induced decoherence. But there is no reason to think that these experiments will not eventually succeed.   

\bigskip

\bigskip

\centerline{\it Newton and Einstein}

\bigskip

\noindent
In 1949, referring to quantum theory, Einstein wrote that it 'would, within the framework of future physics, take an approximately analogous position to that of statistical mechanics within the framework of classical mechanics' \cite{Einstein2}. Not only does this seem to be true as an analogy, but even more precisely so. At the end of the day, Newton reigns supreme. Tables seem to have been turned on quantum theory, and it emerges as the statistical approximation to a classical theory. And then, who would have thought that the phenomenon of Brownian motion, discovered first in the innocuous motion of pollen grains in water, and later brilliantly explained by Einstein as being a consequence of statistical fluctuations, would one day also explain why the macroscopic world around us looks the way it does.    

\bigskip

\bigskip

\noindent {\bf Acknowledgement}: This publication was made possible through the support of a grant
from the John Templeton Foundation. The opinions expressed in this publication are
those of the author and do not necessarily reflect the views of the John Templeton
Foundation.

\newpage

\centerline{\bf Analog Matrix Dynamics : An Example}

\bigskip

\noindent
Given a set of non-commuting spacetime 'coordinates' which are represented by matrices $(\bf{\hat{x}},\hat{t})$ we define an invariant `line-element' as
\begin{equation}
ds^{2} = Tr[d\hat{t}^2 - d\hat{x}^2-d\hat{y}^2-d\hat{z}^2]  = Tr[d\hat{t}'^2 - d\hat{x}'^2 - d\hat{y}'^2 - d\hat{z}'^2]
\end{equation}
where the primed coordinates $({\bf \hat{x}'}, \hat{t}')$ are related to the unprimed ones by a generalized Lorentz transformation. It can be shown that the most general linear transformation which leaves this line-element invariant is of the form
\begin{eqnarray}
 \hat{t}' &=& A \hat{t} + B\hat{x} + \alpha C\hat{y} + \alpha C\hat{z}\nonumber\\
 \hat{x}' &=& A \hat{x} + B\hat{t} + \alpha C\hat{y} + \alpha C\hat{z}\nonumber\\
 \hat{y}' &=&   \hat{y} + D\hat{x} - D\hat{t} \nonumber\\
 \hat{z}' &=&   \hat{z} + D\hat{x} - D\hat{t}, \label{BST}
\end{eqnarray} 
where $\alpha$, $A$ and $B$ are real numbers, $A^2-B^2=1$, $C$ and $D$ are anticommuting Grassmann numbers (whose square is zero by definition), and $D=\alpha (B-A) C$. For a fixed $C$, transformations of the form
$$\left(
\begin{array}{llll}
A & B & \alpha C & \alpha C\\
B & A & \alpha C & \alpha C\\
-D & D & 1 & 0\\
-D & D & 0 & 1
\end{array}
\right)$$
form a group, which is characterized by the fixed $C$. It is straigtforward to show that this line-element possesses, besides invariance under these generalized boosts, invariance also under rotations and translations. In this sense we can associate a generalized Poincare invariance with these noncommuting coordinates. In the limit $\alpha\rightarrow 0$, $D$ goes to zero as well, and the generalized Poincare invariance reduces to the standard Poincare invariance, with each of the coordinates reducing to a c-number times a unit matrix, hence leading to the recovery of ordinary spacetime of commuting coordinates.   

Dynamics is introduced by first defining the four-vector $\hat{x}=({\bf \hat{x}}, t)$, four velocity $\hat{u}^i=d\hat{x}^i/ds$ and four-momentum $\hat{p}^i=m\hat{u}^i$ which satisfies $Tr[\hat{p}^{\mu}\hat{p}_{\mu}]=m^2$. An action operator $\hat{S}$ and the Trace Action $S=\hat{S}$ are introduced :
\begin{equation}
S=Tr\hat{S}=\int d\tau Tr \hat{\cal L}(\hat{x},\dot{\hat{x}})
\end{equation} 
where $\tau$ is the auxiliary parameter (curve parametrization) introduced in the text, and $\dot{\hat{x}}=d\hat{x}/d\tau$. Extremization 
of the action leads to the Lagrange equations for the Trace Lagrangian $L=Tr\hat{\cal L}$
\begin {equation}
 \frac{\partial L}{\partial \hat{x}}-\frac{d}{d\tau}\frac{ \partial L}{ {\partial \dot{\hat{x}}}} = 0.
\end {equation}

Hence,
\begin {equation}
 \frac{ \delta L}{ {\delta \dot{\hat{x}}}} = \int d \tau  \frac{\delta L}{\delta \hat{x}} = \frac{\delta}{\delta \hat{x}}\int d\tau L.    \label{mom}
\end {equation}
If we define $ {\delta L}/{ {\delta \dot{\hat{x}}}}$ as momentum conjugate to $\hat{x}$, then
\begin{equation}
 \hat{P}_x = \frac{\delta S}{\delta \hat{x}}.
\end{equation}
Derivatives with respect to operators are to be understood as Trace derivatives [Adler, 2004].

In classical mechanics, this momentum is tangent to the trajectory in configuration space. Assuming the same to hold here
$ \hat{P}_{x^{\mu}} $ is tangent to the curve drawn in configuration space of $ x^{\mu}$ co-ordinates. Hence,
$ \hat{P}_{x^{\mu}} =\hat{p}^{\mu} $ and
$$ \hat{P}_{\mu} =\frac{\delta S}{\delta \hat{x}^{\mu}} $$
Therefore, the Hamilton-Jacobi equation of motion is 
\begin{equation}
 Tr\left(\left(\frac{\delta S}{\delta \hat{t}}\right)^2 -\left(\frac{\delta S}{\delta \hat{x}}\right)^2 - 
\left(\frac{\delta S}{\delta \hat{y}}\right)^2 - \left(\frac{\delta S}{\delta \hat{z}}\right)^2 
 \right) = m^2.
\end{equation}
The Hamiltonian analog of the dynamics is obtained by constructing the Hamiltonian 
$$ {\cal{H}}= Tr\left( \sum_r \hat{p}_r\dot{\hat{x}}_r -\hat{L} \right) $$
and the Hamilton equations of motion follow by considering the variation
\begin{eqnarray}
 \delta{\cal{H}} &=& Tr \sum_r((\delta\hat{p}_r)\dot{\hat{x}}_r +  \hat{p}_r\delta\dot{\hat{x}}_r) - Tr \sum_r \left(
\frac{\delta L}{\delta \hat{x}_r}\delta \hat{x}_r+\frac{ \delta L}{ {\delta \dot{\hat{x}}_r}}\delta\dot{\hat{x}}\right) \nonumber \\
&=& Tr \sum_r((\delta\hat{p}_r)\dot{\hat{x}}_r -\dot{\hat{p}}_r\delta\hat{x}_r) \nonumber\\
&=& Tr \sum_r(\epsilon_r\dot{\hat{x}}_r\delta\hat{p}_r -\dot{\hat{p}}_r\delta\hat{x}_r)
\end{eqnarray}
where the Trace Hamiltonian is a trace functional of operators ${\hat{x}_r}$, ${\hat{p}_r}$. Thus
$$  \frac{\delta{\cal{H}}}{\delta \hat{x}_r} = -\dot{\hat{p}}_r $$
$$  \frac{\delta{\cal{H}}}{\delta \hat{p}_r} =\epsilon_r\dot{\hat{x}}_r .$$

 If  $\hat{L}$ and ${\cal{H}}$ are constructed using $\hat{x}_r$ and $\dot{\hat{x}}_r$ (or equivalently  $\hat{x}_r$  and ${\hat{p}_r}$) only with 
$c-$ numbers, there is a global unitary invariance which preserves the adjointness property. The charge corresponding to global unitary invariance of
 ${\cal{H}}$: \\
\begin{equation}
 Q = \sum_{r\in B}[\hat{x}_r,\hat{p}_r] -\sum_{r\in F}\{\hat{x}_r,\hat{p}_r\} .
\end{equation}
is the generalized Adler-Millard charge, which is made up of commutators/anti-commutators for bosonic/fermionic degrees of freedom. 

From here, one constructs the statistical thermodynamics of this Analog Dynamics, thereby obtaining a quantum-theory like digital description for 
matrix-valued matter and spacetime degrees of freedom. Consideration of Brownian motion fluctuations leads to the classical approximation for matter fields and spacetime.

Details will appear in Lochan and Singh (2011, in preparation).

\end{document}